\documentclass{elsart}
\usepackage{natbib}

\usepackage{amssymb}
\usepackage{graphicx}

\newcommand{\vecbm}[1]{\mbox{\boldmath#1}}
\newcommand{\vecb}[1]{\mbox{\bf#1}}
\newcommand{\cent}[1] {\begin{center}#1\end{center}}

\begin{document}
\begin{frontmatter}
\title{Negative heat-capacity at phase-separations in microcanonical thermostatistics of macroscopic
systems\\ with either short or long-range interactions}
\author{D.H.E. Gross}
\address{Hahn-Meitner Institute and Freie Universit{\"a}t Berlin,\\
Fachbereich Physik.\\ Glienickerstr. 100\\ 14109 Berlin, Germany\\
gross@hmi.de; http://www.hmi.de/people/gross}

\begin{abstract}
Conventional thermo-statistics address infinite homogeneous systems within
the canonical ensemble. However, some 170 years ago the original motivation
of thermodynamics was the description of steam engines, i.e. boiling water.
Its essential physics is the separation of the gas phase from the liquid.
Of course, boiling water is inhomogeneous and as such cannot be treated by
conventional thermo-statistics. Then it is not astonishing, that a phase
transition of first order is signaled canonically by a Yang-Lee
singularity. Thus it is only treated correctly by microcanonical
Boltzmann-Planck statistics. This was elaborated in the talk presented at
this conference. It turns out that the Boltzmann-Planck statistics is much
richer and gives fundamental insight into statistical mechanics and
especially into entropy. This can be done to a far extend rigorously and
analytically. The deep and essential difference between ``extensive'' and
``intensive'' control parameters, i.e. microcanonical and canonical
statistics, was exemplified by rotating, self-gravitating systems. In the
present paper the necessary appearance of a convex entropy $S(E)$ and the
negative heat capacity at phase separation in small as well macroscopic
systems independently of the range of the force is pointed out.
\end{abstract}

\end{frontmatter}
\section{Introduction}
Since the beginning of thermodynamics in the first half of the
19.century its original motivation was the description of steam engines and
the liquid to gas transition of water. Here water becomes inhomogeneous and
develops a separation of the gas phase from the liquid, i.e. water boils.

A little later statistical mechanics was developed by
Boltzmann\cite{boltzmann1872} to explain the microscopic mechanical basis
of thermodynamics. Up to now it is generally believed that this is given by
the Boltzmann-Gibbs canonical statistics. As traditional canonical
statistics works only for homogeneous, infinite systems, phase separations
remain outside of standard Boltzmann-Gibbs thermo-statistics, which,
consequently, signal phase-transitions of first order by Yang-Lee
singularities.

It is amusing that this fact that is essential for the original purpose of
thermodynamics to describe steam engines was never treated completely in
the past 150 years.  The system must be somewhat artificially split into
(still macroscopic and homogeneous) pieces of each individual phase
\cite{guggenheim67}. The most interesting configurations of two coexisting
phases cannot be described by a single canonical ensemble. Important
inter-phase fluctuations remain outside, etc. This is all hidden due to the
restriction to homogeneous systems in the thermodynamic limit.

Also the second law can rigorously be formulated only microcanonically:
Already Clausius \cite{clausius1854} distinguished between external and
internal entropy generating mechanisms. The second law is only related to
the latter mechanism \cite{prigogine71}, the internal entropy generation.
Again, canonical Boltzmann-Gibbs statistics is insensitive to this
important difference.

For this purpose, and also to describe small systems like fragmenting
nuclei or non-extensive ones like macroscopic systems at phase-separation,
or even very large, self-gravitating, systems, we need a new and deeper
definition of statistical mechanics and as the heart of it: of entropy. For
this purpose it is crucial to avoid the thermodynamic limit.

As the main aspects of this new thermodynamics were published in
\cite{gross216,gross215,gross186,gross213,gross214} I skip here to repeat
all the arguments. Instead I will stress here only the fact that negative
heat capacity and convex entropy can be seen at proper phase transitions of
1. order, i.e. at phase {\em separation}, in small as well in macroscopic
systems independently whether they have long or short range interactions.
As there was a hot discussion at this conference about this point, it seems
necessary to repeat the arguments here.
\section{Negative heat capacity at phase-separation can also be seen in macroscopic systems
independently of the range of the interaction. } The argument is simple
c.f.\cite{gross214}: At phase separation the weight $e^{S(E)-E/T}$ of the
configurations with energy E in the canonical partition sum
\begin{equation}
Z(T)=\int_0^\infty{e^{S(E)-E/T}dE}\label{canonicweight}
\end{equation} becomes {\em bimodal}, at the transition temperature it has
two peaks, the ``liquid'' and the ``gas'' configurations which are
separated in energy by the latent heat. Consequently $S(E)$ must be convex
($\partial^2 S/\partial E^2>0$, like $y=x^2$) and the weight in
(\ref{canonicweight}) has a minimum at $E_{min}$ between the two pure
phases. Of course, the minimum can only be seen in the microcanonical
ensemble where the energy is controlled and its fluctuations forbidden.
Otherwise, the system would fluctuate between the two pure phases by an,
for macroscopic systems even macroscopic, energy $\Delta E\sim
E_{lat}\propto N$ of the order of the latent heat in clear contrast to the
usual assumption of the fluctuations in the canonical ensemble $\delta
E\propto \sqrt{N}$ . The heat capacity is
\begin{equation}
C_V(E_{min})=\partial E/\partial T=-~~\left.\frac{(\partial S/\partial
E)^2}{\partial^2S/\partial E^2}\right|_{E_{min}}<0.\label{negheat}
\end{equation}
I.e. {\em the convexity of $S(E)$ and the negative microcanonical heat
capacity are the generic and necessary signals of  any
phase-separation\cite{gross174}}.

This ``convex intruder'' in $S(E)$ with the depth $\Delta
S_{surf}(E_{min})$ has a direct physical significance: Its depth is the
surface entropy due to constraints by the existence of the inter-phase
boundary between the droplets of the condensed phase and the gas phase and
the corresponding correlation. $\Delta S_{surf}(E_{min})$ is directly
related to the surface tension per surface atom (with number $N_{surf}$) of
the droplets.
\begin{equation}
\sigma_{surf}/T_{tr}=\frac{\Delta S_{surf}(E_{min})}{N_{surf}}
\end{equation}

In my paper together with M.Madjet \cite{gross157} we have compared the
values of $\Delta S_{surf}(E_{min})$ calculated by Monte-Carlo using a
realistic short range interaction with the values of the surface tension of
the corresponding macroscopic system. In these calculations we used the
empirical liquid drop parameters for the  ground-states energies of the
different clusters as given by \cite{brechignac95}.
\begin{figure}[h]\cent{
\includegraphics*[bb = 99 57 400 286, angle=-0, width=8cm,
clip=true]{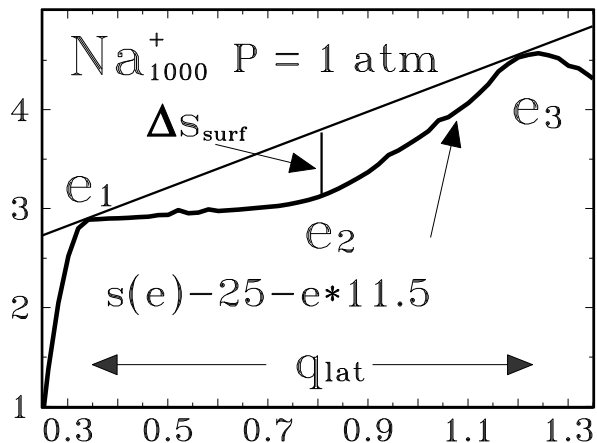}} \caption{Microcanonical Monte-Carlo
(MMMC)~\protect\cite{gross157,gross174} simulation of the entropy
  \index{entropy} $s(e)$ per atom ($e$ in eV per atom) of a system of
  $N=1000$ sodium atoms at an external pressure of 1 atm.  At the
  energy $e\leq e_1$ the system is in the pure liquid phase and at
  $e\geq e_3$ in the pure gas phase, of course with fluctuations. The
  latent heat per atom is $q_{lat}=e_3-e_1$.  \underline{Attention:}
  the curve $s(e)$ is artificially sheared by subtracting a linear
  function $25+e*11.5$ in order to make the convex intruder visible.
  {\em $s(e)$ is always a steep monotonic rising function}.  We
  clearly see the global concave (downwards bending) nature of $s(e)$
  and its convex intruder. Its depth is the entropy loss due to
  additional correlations by the interfaces. It scales $\propto
  N^{-1/3}$. From this one can calculate the surface
  tension per surface atom
  $\sigma_{surf}/T_{tr}=\Delta s_{surf}*N/N_{surf}$.  The double
  tangent (Gibbs construction) is the concave hull of $s(e)$. Its
  derivative gives the Maxwell line in the caloric curve $e(T)$ at
  $T_{tr}$. In the thermodynamic limit the intruder in $s(e)$ would disappear
  and $s(e)$ would approach the double tangent from below, not of course in
  $S(E)$, which remains deeply convex: The probability of configurations with
  phase-separations is suppressed by the (infinitesimal small)
  factor $e^{-N^{2/3}}$ relative to the pure phases and the
  distribution remains {\em strictly bimodal in the canonical ensemble}
  in which the region $e_1<e<e_3$ of phase separation gets lost.\label{naprl0f}}
\end{figure}
Table (\ref{table}) shows the scaling behavior of $\Delta
S_{surf}(E_{min})$ with the size $N$ of the system.
\begin{table}
\caption{Parameters of the liquid--gas transition of small
  sodium clusters (MMMC-calculation~\protect\cite{gross157,gross174}) in
  comparison with the bulk for a rising number $N$ of atoms,
  $N_{surf}$ is the average number of surface atoms (estimated here as
  $\sum{N_{cluster}^{2/3}}$) of all clusters with $N_i\geq2$ together.
  $\sigma/T_{tr}=\Delta s_{surf}*N/N_{surf}$ corresponds to the
  surface tension. Its bulk value is adjusted to agree with the
  experimental values of the $a_s$ parameter which
  we used in the liquid-drop formula for the binding energies of small
  clusters, c.f.  Brechignac et al.~\protect\cite{brechignac95}, and
  which are used in this calculation~\cite{gross174} for the individual
  clusters.\label{table}}
\begin{center}
\renewcommand{\arraystretch}{1.4}
\setlength\tabcolsep{5pt}
\begin{tabular} {|c|c|c|c|c|c|} \hline
&$N$&$200$&$1000$&$3000$&\vecb{bulk}\\ 
\hline \hline &$T_{tr} \;[K]$&$940$&$990$&$1095$&\vecb{1156}\\ \cline{2-6}
&$q_{lat} \;[eV]$&$0.82$&$0.91$&$0.94$&\vecb{0.923}\\ \cline{2-6} {\bf
Na}&$s_{boil}$&$10.1$&$10.7$&$9.9$&\vecb{9.267}\\ \cline{2-6} &$\Delta
s_{surf}$&$0.55$&$0.56$&$0.44$&\\ \cline{2-6}
&$N_{surf}$&$39.94$&$98.53$&$186.6$&\vecbm{$\infty$}\\ \cline{2-6}
&$\sigma/T_{tr}$&$2.75$&$5.68$&$7.07$&\vecb{7.41}\\ \hline
\end{tabular}
\end{center}
\end{table}
\newpage
Roughly $\Delta S_{surf}(E)\propto N^{2/3}$ and one may argue that this
will vanish compared to the ordinary leading volume term $S_{vol}(E)\propto
N$. However, this is not so as $S_{vol}(E)$ at energies inside the
phase-separation region (the convex intruder) is the \underline{concave
hull} of $S(E)$ (its slope gives the Maxwell construction of the caloric
curve $T(E)$). It is a straight line and its curvature $\partial^2
S_{vol}/\partial E^2\equiv 0$. Consequently for large $N$
\begin{equation}
\partial^2S/\partial E^2\sim
\partial^2 S_{vol}/\partial E^2+\partial^2\Delta S_{surf}/\partial E^2
+\cdots \asymp \partial^2\Delta S_{surf}/\partial E^2
\end{equation} and the depth of the intruder $\Delta S_{surf}(E_{min})=\sigma/T_{tr}
*N_{surf}\sim N^{2/3}$ goes to infinity in the thermodynamic limit. Of
course, the ubiquitous phenomena of phase separation exist only by this
reason. It determines the (negative) heat capacity as in
eq.(\ref{negheat}). The physical (quite surprising) consequences are
discussed in \cite{gross214,gross213}.

Discussions with St.Ruffo are gratefully acknowledged.


\end{document}